\documentclass[onecolumn,showpacs,nobibnotes,nofootinbib,12pt]{revtex4}
\usepackage{amsmath,amssymb}
\usepackage{wasysym}
\usepackage{slashed}
\usepackage{graphicx}
\usepackage{pdfcomment}
\newcommand{\be}{\begin{equation}}
\newcommand{\ee}{\end{equation}}
\newcommand{\bea}{\begin{eqnarray}}
\newcommand{\eea}{\end{eqnarray}}
\newcommand{\f}{\frac}

\newcommand{\nn}{\nonumber}

\newcommand{\bra}{\langle}
\newcommand{\ket}{\rangle}
\newcommand{\D}{\triangledown}

\begin{document}
\title{Radiative energy loss and radiative $p_\perp$-broadening of high-energy partons in QCD matter}
\author{Bin Wu}
\email{Bin.WU@cea.fr}
\affiliation{Institut de Physique Th\'{e}orique, CEA Saclay, 91191, Gif-sur-Yvette Cedex, France}
\begin{abstract}
We study the connection between radiative energy loss and radiative $p_\perp$-broadening of a high-energy quark or gluon passing through QCD matter. The generalized Baier-Dokshitzer-Mueller-Peigne-Schiff-Zakharov (BDMPS-Z) formalism is used to calculate energy loss due to multiple gluon emission. With $L$ the length of the matter and $l_0$ the size of constituents of the matter we find a double logarithmic correction $\propto\ln^2\f{L}{l_0}$ to parton energy loss due to two-gluon emission. We also show that the radiative energy loss per unit length $-\f{dE}{dz} = \f{\alpha_s N_c}{12} \bra p_\perp^2\ket$ by carrying out a resummation of the double logarithmic terms. Here, the transverse momentum broadening $\bra p_\perp^2\ket$ is obtained by resumming terms $\propto\alpha_s \ln^2\f{L}{l_0}$ in Ref. \cite{Liou:Mueller:Wu:2013}. Our result agrees with that by the renormalization of $\hat{q}$ proposed in Refs. \cite{Blaizot:Mehtar-Tani:2014,Iancu:2014}. 
\end{abstract}
\maketitle
%
%
\section{Introduction}
Jet quenching is an important signal of the production of dense QCD matter in the ultra-relativistic heavy-ion collisions at RHIC\cite{Adler:2003,Adams:2003} and LHC\cite{ALICE:2010,CMS:2012,ATLAS:2012}. Parton energy loss in QCD matter is believed to be responsible for such a suppression of the yields of large $p_\perp$ hadrons and jets with respect to $p+p$ collisions (see \cite{Muller:2012} for a recent review). 
  
The radiative energy loss and the $p_\perp$-broadening of a high-energy parton in QCD matter are closely related to each other. Energy loss is dominated by radiating a gluon with the maximum energy $\omega_c$. The gluon picks up a transverse momentum broadening $\bra p_\perp^2\ket = \f{\omega_c}{t_c}$ within the coherent (formation) time $t_c$. As a result, the energy loss of the parton per unit length is given by\cite{BDMPS:1996pt}
\be\label{eq:dept2}
-\f{dE}{dz} \sim \alpha_s N_c \f{\omega_c}{t_c} = \alpha_s N_c \bra p_\perp^2\ket.
\ee
In a medium of length $L$ the $p_\perp$-broadening due to multiple scattering is given by $\bra p_\perp^2\ket = \hat{q} L$ with $\hat{q}$ being the transport coefficient. If one only considers multiple scattering, radiative energy loss is dominated by one gluon with  $t_c \simeq L$ and $\omega_c\simeq\hat{q} L^2$. In this case one has\cite{BDMPS:1996,BDMPS:1998}
\be
-\f{dE}{dz} \sim \alpha_s N_c \hat{q} L.
\ee

The radiative $p_\perp$-broadening of a high-energy parton in QCD matter is first studied in Ref. \cite{Wu:2011}. Double logarithmic terms due to the recoil effect of one-gluon emission are found in the kinetic region of single scattering. The complete result of such a double logarithmic correction is obtained in Ref. \cite{Liou:Mueller:Wu:2013}, which takes the form 
\be\label{eq:pt2rad}
\bra p_\perp^2 \ket_{rad} = \f{\alpha_s N_c}{8 \pi}\hat{q} L\ln^2\left(\f{L}{l_0}\right)^2,
\ee
where $l_0$ is the size of constituents of the matter. Moreover, the resummation of the double logarithmic terms can be carried out to give\cite{Liou:Mueller:Wu:2013}
\be\label{eq:pt2}
\bra p_\perp^2 \ket=\hat{q} L\sqrt{\f{4\pi}{\alpha_s N_c}}\f{1}{\ln \f{L^2}{l_0^2}}I_1\left[\sqrt{\f{\alpha_s N_c}{\pi}}  \ln \f{L^2}{l_0^2} \right].
\ee

The transport coefficient $\hat{q}$ can be written as the expectation value of a gauge-invariant operator, which is  proportional to the gluon distribution function of the medium\cite{BDMPS:1996pt} and can be studied non-perturbatively via
simulations on a Euclidean lattice\cite{Panero:2013}. A renormalization of $\hat{q}$, based on the DGLAP evolution of the gluon distribution, has been proposed in Ref. \cite{CasalderreySolana:Wang:2007} (see \cite{Xing:2014} for a recent development of such a proposal). More recently, Refs. \cite{Blaizot:Mehtar-Tani:2014, Iancu:2014} propose another evolution equation for the renormalization of $\hat{q}$, valid in the double logarithmic approximation. This equation applies in the regime of multiple soft scattering while the former one in Ref. \cite{CasalderreySolana:Wang:2007} is better suited for the study of the high-momentum tail of the $p_\perp$-broadening associated with a single hard scattering\cite{Iancu:2014}. The solution $\hat q_{\rm ren}(L)$ to the equation in \cite{Blaizot:Mehtar-Tani:2014, Iancu:2014} is consistent with the result in eq. (\ref{eq:pt2}) provided it is rewritten as $\bra p_\perp^2\ket={\hat q_{\rm ren}}L$.  Based on such a proposal, eq. (\ref{eq:dept2}) is expected to hold for radiating an arbitrary number of gluons in the double logarithmic approximation.

In this paper we give a detailed calculation of the double logarithmic correction to the energy loss of a high-energy parton by radiating two or more gluons in QCD matter. Our aim is to go beyond the parametric estimate in eq. (\ref{eq:dept2}) and to show explicitly how the double logarithmic correction to radiative energy loss is related to the radiative $p_\perp$-broadening in eq. (\ref{eq:pt2rad}). 

The paper is organized as follows. In Sec. \ref{sec:BDMPSZ}, we first give the general formalism, as a generalization of that by BDMPS-Z\cite{BDMPS:1996,BDMPS:1998,Zakharov:1996,Zakharov:1997}, for calculating parton energy loss due to multiple gluon emission. Then, we give all the diagrams relevant for energy loss due to two-gluon emission. The time-evolution of two gluons together with a dipole in the medium is studied in Sec. \ref{sec:evolution}. In Sec. \ref{sec:doublelog} we evaluate the double logarithmic correction to radiative energy loss in details. Our conclusion is presented in Sec. \ref{sec:conclusion}.

\section{Medium-induced energy loss due to two-gluon emission}\label{sec:BDMPSZ}
\subsection{Review of the generalized BDMPS-Z formalism for radiative energy loss}
In this subsection we first give the general formalism for medium-induced energy loss of a high-energy parton due to multiple gluon emission. The formalism, used to calculate the radiative $p_\perp$-broadening in Refs.\cite{Wu:2011, Liou:Mueller:Wu:2013}, is a slight extension of the BDMPS-Z formalism\cite{BDMPS:1996,BDMPS:1998, Zakharov:1996, Zakharov:1997}\footnote{The interested reader is referred to Refs. \cite{Kovner:Wiedemann:2003,Majumder:2014} and references therein for different approaches used to calculate parton energy loss.} by including virtual gluon emission. In this paper we choose to use the path integral representation in the transverse coordinate space\cite{Zakharov:1996,Zakharov:1997, Liou:Mueller:Wu:2013}. 

To calculate the medium-induced energy loss of a high-energy parton, one needs the spectrum of $n$ soft gluons. Since the gluons can be real or virtual, the spectrum shall be denoted by 
\be
\int d\omega_{m+1}d\omega_{m+2}\cdots d\omega_n \f{d I^{(m,n-m)} }{d\omega_1 d\omega_2\cdots d\omega_n}
\ee
where the energyies of $m$ ($\leq n$) real gluons are respectively denoted by $\omega_1, \omega_2, \cdots, \omega_m$ and the energies of $(n-m)$ virtual gluons are respectively denoted by $\omega_{m+1}, \omega_{m+2}, \cdots, \omega_n$. The spectrum can be calculated using the following steps\footnote{In this paper, the calculation is done using lightcone gauge in ordinary space-time coordinates\cite{Wu:2011,Liou:Mueller:Wu:2013}. One can carry out the same calculation using lightcone gauge in lightcone coordinates (see, e.g., \cite{Iancu:2014}). In this case, there are additional contributions from instantaneous Coulomb terms of the background propagators\cite{Iancu:2014}, which, however, do not contribute to the double logarithmic correction to parton energy loss.}
\begin{enumerate}
\item Draw all the relevant graphs with $m$ real gluons and $(n-m)$ virtual gluons.\\
Inside the medium it is usually more convenient to draw graphs as a product of amplitudes and their complex conjugates.
\item Obtain the contributions of each graph from the following Feynman rules
\bea
&&\begin{array}{c}\includegraphics[width=3cm]{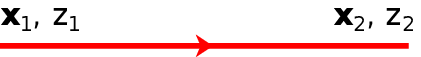}\end{array}= \int\limits_{{\bf x}(z_1) = {\bf x}_1}^{{\bf x}(z_2) = {\bf x}_2} D{\bf{x}} e^{i\f{\omega}{2}\int_{z_1}^{z_2}dt~\dot{\bf x}^2}P e^{ ig\int_{z_1}^{z_2}dt A^{a-}(t, {\bf  x}(t))T_R^a},\\
&&\begin{array}{c}\includegraphics[width=3cm]{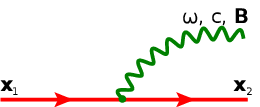}\end{array}=\delta({\bf B}-{\bf x}_1) \delta({\bf x}_2-{\bf x}_1)\f{g}{\omega} T^c_R \epsilon^*_{\perp\lambda}\cdot \nabla_{\bf B},
\eea
where $T^a_R$ is the $SU(N_c)$ matrix in the representation $R$ corresponding to the high-energy parton, which can be either a quark ($R=F$) or a gluon ($R=A$) represented by the solid lines in the above graphs, and the gluon transverse polarization vector $\epsilon_{\lambda\perp}$ satisfies
\be
\sum\limits_{\lambda}\epsilon_{\lambda\perp}^i \epsilon_{\lambda\perp}^{*j} = \delta_{ij}.
\ee
In this paper vectors in the transverse plane are denoted by bold letters.
\item Put in the overall prefactor $\f{1}{(4\pi)^n}$.
\item Integrate out the background medium.\\
The background medium is modelled by the background gluon field $A^{a-}(t,{\bf x})$ with $a$ the color index. The ensemble average over the background field is defined by
\be
\bra A^{a-}(t_1, {\bf{x}}) A^{b-}(t_2,{\bf{y}}) \ket= \delta(t_1-t_2) \delta_{ab} n(t_1) \Gamma({\bf{x}-\bf{y}}),
\ee 
which is related to the transport coefficient $\hat{q}$ by
\be
g^2 C_R n(t) \left[ \Gamma({\bf{0}}) - \Gamma({\bf{x}}) \right]\equiv n(t)\left[ \sigma_R({\bf{0}}) - \sigma_R({\bf{x}}) \right] \simeq \f{1}{4} \hat{q} x_\perp^2
\ee 
where $n(t)$ is the number density of the scatterers and $\sigma_R({\bf x})$ is defined by\cite{Wu:2011}
\be
\sigma_R({\bf x})\equiv\int\f{d^2{\bf q}}{(2\pi)^2} e^{i{\bf q\cdot x}}\f{d\sigma_R}{d^2{\bf q}}=\f{\alpha_s C_R}{\pi}\int\f{d^2{\bf q}}{(2\pi)^2} e^{i{\bf q\cdot x}}|A^-(t,{\bf q})|^2
\ee
with $C_R = N_c$ for gluons and $C_R = C_F\equiv\f{N_c^2-1}{2N_c}$ for quarks.

\end{enumerate}
\subsection{Diagrams for the energy loss due to two-gluon emission}
\begin{figure}
\begin{center}
\includegraphics[width=0.8\textwidth]{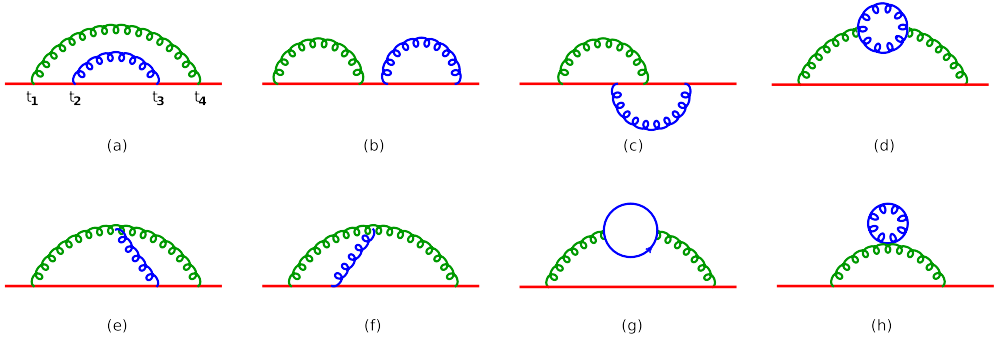}
\end{center}
\caption{Forward scattering amplitudes at $O(\alpha_s^2)$. All the relevant graphs for calculating the energy loss due to two-gluon emission can be obtained by cutting through these amplitudes.}\label{fig:all}
\end{figure}
In this subsection we give all the graphs for the energy loss due to two-gluon emission, which is denoted by $\Delta E_2$. In these graphs at least one of the two gluons is real. In terms of the spectra defined in the previous subsection, $\Delta E_2$ is given by
\be
\Delta E_{2} = \int d\omega_1 d\omega_2 \left( \omega_1 + \omega_2 \right)  \f{d I^{(2,0)}}{d\omega_1 d\omega_2} + \int d\omega_1 d\omega_2 \omega_1 \f{dI^{(1,1)}}{d\omega_1 d\omega_2}.
\ee
All the graphs contributing to $\Delta E_2$ can be obtained by cutting through the forward amplitudes in Fig. \ref{fig:all}. The cuts of Fig. \ref{fig:all} (h) are not relevant for medium-induced energy loss and shall be ignored. Besides, we shall also ignore the graphs obtained by cutting Fig. \ref{fig:all} (g) because the gluon-quark transition is suppressed compared to the soft gluon emission at high energies. Then, we are left with 17 possible cuts of Fig. \ref{fig:all} (a) to (f).

Even in the same graph, different orderings in the 4 emission (absorption) times of the two gluons give different contributions, which need to be dealt with separately. The complete calculate of $\Delta E_2$ involves all the possible orderings in these 4 time variables. The graph with one of such orderings is referred to as a diagram in this paper. It is easy to show that there are a total of $78$ different diagrams\footnote{There are 6 possible orderings in the 4 time variables for a graph with two real gluons while there are only 4 for a graph with only one real gluon. Since there are respectively 5 graphs with two real gluons and 12 graphs with only one real gluon, one has 78 orderings in total.}, which can be classified as follows 

\begin{itemize}
\item{\bf 12 uncorrelated emissions:}\\
In these diagrams both of the two emission times of one gluon are later than those of the other. The distribution of the uncorrelated soft gluons is the QCD analog of that of the soft photons in QED, which has the form of a Poisson distribution after all the uncorrelated multiple gluon emissions are included\cite{BDMS:2001:uncorrelated}. And there is no double logarithmic correction proportional to $\ln^2\left(\f{L}{l_0}\right)^2$ from these diagrams. 
\item{\bf 26 fully-overlapping emissions:}\\
In these diagrams both of the two emission times of one gluon lie in between those of the other gluon. In the following sections  we shall show that there is a double logarithmic correction to $\Delta E_2$ from these diagrams, which is the same as that to $\bra p_\perp^2\ket$ in Ref. \cite{Liou:Mueller:Wu:2013}.
\item{\bf 40 partially-overlapping emissions:}\\
In these diagrams only one of the two emission times of one gluon lies in between those of the other gluon. The evaluation of those diagrams is the most difficult part to obtain the complete result of $\Delta E_2$. Fortunately, unlike the fully-overlapping emissions the diagrams contributing to the double logarithmic terms $\propto\ln^2\f{L}{l_0}$ of $\bra p_\perp^2 \ket$ do not show up as subdiagrams of these diagrams. Therefore, there is no double logarithmic correction the same as that in $\bra p_\perp^2\ket$ from these diagrams.
\end{itemize}
\section{The time evolution of two gluons in the medium}\label{sec:evolution}
\begin{figure}  
\begin{center}
\includegraphics[width=0.5\textwidth]{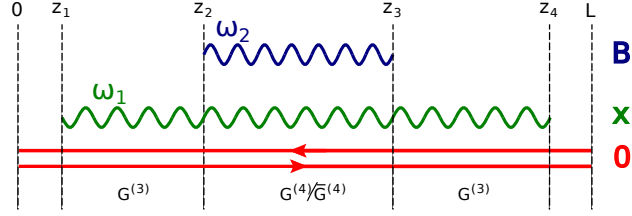}
\end{center}
\caption{Time-evolution of two gluons and a dipole in the medium. The first gluon with energy $\omega_1$ emitted off the quark at $t=z_1$ lives until $t=z_4$. The second gluon with energy $\omega_2$ is emitted at $t=z_2$ and absorbed at $t=z_3$ either by the first gluon or by the dipole. The transverse coordinates of the first and the second gluon are respectively denoted by ${\bf x}$ and ${\bf B}$, and the dipole locates at ${\bf{0}}$. Here, the time variables satisfy $0<z_1<z_2<z_3<z_4<L$.}\label{fig:dipole}
\end{figure}
In this paper we use a dipole-like picture\footnote{ Here, the dipole can be either a quark-antiquark pair or a gluon pair, representing the same high-energy parton in the amplitude and in the conjugate amplitude. Our results are valid both for high-energy quarks and gluons by choosing $\hat{q}$ accordingly.} to describe the whole process of two-gluon emission in the QCD medium\cite{Liou:Mueller:Wu:2013}. In our calculation we shall only include the fully-overlapping emissions  as illustrated in Fig. \ref{fig:dipole}:  the first gluon with energy $\omega_1$ emitted off the quark at $t=z_1$ lives until $t=z_4$; and the second gluon with energy $\omega_2$ is emitted at $t=z_2>z_1$ and absorbed at $t=z_3<z_4$ either by the first gluon or by the dipole. In our case the energy of the high-energy parton $E\gg \hat{q} L^2$ and, therefore, the change of the transverse coordinates of the dipole can be neglected. Due to the homogeneity of the medium in the transverse plane, our result should be independent of the transverse coordinates of the dipole, which are chozen to be ${\bf 0}$.

The time evolution of one gluon with energy $\omega_1$ together with the dipole inside the medium is described by $G^{(3)}$\cite{BDMPS:1996,BDMPS:1998,Zakharov:1996, Zakharov:1997}
\bea\label{eq:G3}
&&G^{(3)}({\bf B}_2, z_2, {\bf B}_1, z_1; \omega_1)\nn\\
&&\equiv\int\limits_{{\bf B}(z_1) = {\bf B}_1}^{{\bf B}(z_2) = {\bf B}_2} D{\bf B}\exp\left\{\int_{z_1}^{z_2} d\xi \left[ i\f{\omega_1}{2}\dot{{\bf B}}^2 +  \f{N_c n(\xi)}{C_R} \left( \sigma_R({\bf B}, \xi) - \sigma_R(0_\perp, \xi) \right) \right] \right\}.
\eea
For the medium-induced energy loss the harmonic oscillator approximation is justified\cite{BDMPS:1996} and one has
\be\label{eq:G3sp}
G^{(3)}({\bf B}_2, z_2, {\bf B}_1, z_1; \omega) \simeq G({\bf B}_2, {\bf B}_1, z_2-z_1; \omega_1, \Omega_1),
\ee
where
\be
\Omega_j = \f{1-i}{2}\sqrt{\f{\hat{q}_A}{\omega_j}}\qquad\text{ with } j = 1,2\text{ and }
\hat{q}_A\equiv \f{N_c}{C_R} \hat{q},
\ee 
and the propagator of the harmonic oscillator
\be
G({\bf B}_2, {\bf B}_1, z, \omega, \Omega)\equiv \f{\omega \Omega}{2 \pi i \sin(\Omega z)} \exp\left\{ \f{i \omega \Omega}{2 \sin(\Omega z)}\left[ \left( {\bf B}_2^2 + {\bf B}_1^2 \right)\cos(\Omega z) - 2 {\bf B}_2\cdot {\bf B}_1 \right] \right\}.
\ee

\begin{figure}
\begin{center}
\includegraphics[width=0.7\textwidth]{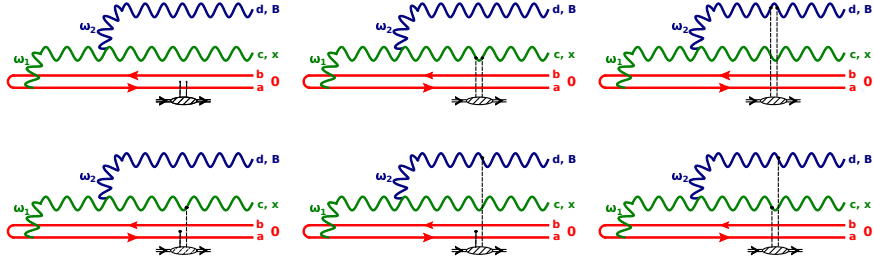}
\end{center}
\caption{Scattering of two gluons and a dipole off one scatterer. All the possible hookings of the two gluon lines issued from the scatterer on the 4 partons are shown in this figure. In the cases when one gluon line of the scatterer is hooked either on the quark line or on the antiquark line of the dipole, the end of the gluon line is put in between the quark-antiquark pair. Here, $a, b, c$ and $d$ are color indices, ${\bf 0}, {\bf x}$ and ${\bf B}$ are the transverse coordinates, and $\omega_1$ and $\omega_2$ are the energies of the two gluons.}\label{fig:G4}
\end{figure}

During $z_3>t>z_2$ one needs to understand how the two gluons together with the dipole evolve in the medium. Fig. \ref{fig:G4} shows all the possible diagrams for the scattering of the two gluons and the dipole off one scatterer. Without any scattering one has the color matrix $(T_A^d)_{c e} T^e_R$ for the 4-body system. Including the scattering off one scatterer one has $\f{N_c}{2}(T_A^d)_{c e} T^e_R$ for each diagram in the figure, which gives the color factor $\f{N_c}{2}$ for each diagram. As a result the potential of the evolution Hamiltonian of the 4-body system can be easily calculated, which takes the form
\be
-iV^{(4)}(t, {\bf x}, {\bf B}) =  \f{N_c n(t)}{2 C_R} \left[ \sigma_R({\bf x}) +  \sigma_R({\bf B}) +  \sigma_R({\bf B}-{\bf x}) - 3 \sigma_R({\bf 0}) \right].
\ee
Let us assume that the first gluon is emitted off the quark of the dipole. If the second gluon is emitted off the first gluon or the quark, the propagator of the 4-body system is given by
\bea\label{eq:G4}
&&G^{(4)}({\bf x}_3, {\bf B}_3,  z_3; {\bf x}_2, {\bf B}_2, z_2;\omega_1,\omega_2)\nn\\
&&\equiv\int\limits_{{\bf x}(z_2) = {\bf x}_2, {\bf B}(z_2) = {\bf B}_2}^{{\bf x}(z_3) = {\bf x}_3, {\bf B}(z_3) = {\bf B}_3} D{\bf x} D{\bf B} e^{i\int_{z_1}^{z_2} d\xi \left[ \f{1}{2}\omega_1 \dot{{\bf x}}^2+\f{1}{2}\omega_2\dot{{\bf B}}^2-V^{(4)}(\xi, {\bf x}, {\bf B})\right]}.
\eea
Otherwise, the propagator takes the form
\bea\label{eq:G4bar}
&&\bar{G}^{(4)}({\bf x}_3, {\bf B}_3,  z_3; {\bf x}_2, {\bf B}_2, z_2;\omega_1,\omega_2)\nn\\
&&\equiv\int\limits_{{\bf x}(z_2) = {\bf x}_2, {\bf B}(z_2) = {\bf B}_2}^{{\bf x}(z_3) = {\bf x}_3, {\bf B}(z_3) = {\bf B}_3} D{\bf x} D{\bf B} e^{i\int_{z_1}^{z_2} d\xi \left[ \f{1}{2}\omega_1 \dot{{\bf x}}^2-\f{1}{2}\omega_2\dot{{\bf B}}^2-V^{(4)}(\xi, {\bf x}, {\bf B})\right]}\nn\\
&&=G^{(4)}({\bf x}_3, {\bf B}_3,  z_3; {\bf x}_2, {\bf B}_2, z_2;\omega_1,-\omega_2).
\eea
In the harmonic oscillator approximation one has
\bea\label{eq:G4sp}
&&G^{(4)}({\bf x}_3, {\bf B}_3,  z_3; {\bf x}_2, {\bf B}_2, z_2;\omega_1,\omega_2)\nn\\
&&\simeq\int\limits_{{\bf x}(z_2) = {\bf x}_2, {\bf B}(z_2) = {\bf B}_2}^{{\bf x}(z_3) = {\bf x}_3, {\bf B}(z_3) = {\bf B}_3} D{\bf x} D{\bf B} e^{\int_{z_1}^{z_2} d\xi \left\{ \f{i}{2}\omega_1 \dot{{\bf x}}^2+\f{i}{2}\omega_2\dot{{\bf B}}^2 -\f{\hat{q}_A}{8} \left[ {\bf x}^2 +  {\bf B}^2 +  ({\bf B}-{\bf x})^2\right]\right\}}\nn\\
&&= G(\tilde{{\bf x}}_3, \tilde{{\bf x}}_2, z_3 - z_2, m_1, K_1) G(\tilde{{\bf B}}_3, \tilde{{\bf B}}_2, z_3 - z_2, m_2, K_2),
\eea
where the new coordinates are defined as
\bea
\left(\begin{array}{c}
\tilde{{\bf x}}\\
\tilde{{\bf B}}
\end{array}
\right)=\left(
\begin{array}{cc}
 1 & \frac{-\omega_1+\omega_2+\sqrt{\omega_1^2-\omega_1 \omega_2+\omega_2^2}}{\omega_1} \\
 -\frac{1}{2} & \frac{\omega_1-\omega_2+\sqrt{\omega_1^2-\omega_1 \omega_2+\omega_2^2}}{2 \omega_1} \\
\end{array}
\right)
\left(\begin{array}{c}
{\bf x}\\
{\bf B}
\end{array}
\right),
\eea
and
\bea
&&m_1=\frac{\omega_1 \left(\omega_1-\omega_2+\sqrt{\omega_1^2-\omega_1 \omega_2+\omega_2^2}\right)}{2 \sqrt{\omega_1^2-\omega_2 \omega_1+\omega_2^2}},\\
&&m_2=\frac{2 \omega_1 \left(\omega_2-\omega_1+\sqrt{\omega_1^2-\omega_1 \omega_2+\omega_2^2}\right)}{\sqrt{\omega_1^2-\omega_2 \omega_1+\omega_2^2}},\\
&&K_1^2=-\frac{i \hat{q}_A \left(\omega_1+\omega_2-\sqrt{\omega_1^2-\omega_1 \omega_2+\omega_2^2}\right)}{4 \omega_1 \omega_2},\\
&&K_2^2=-\frac{i \hat{q}_A \left(\omega_1+\omega_2+\sqrt{\omega_1^2-\omega_1 \omega_2+\omega_2^2}\right)}{4 \omega_1 \omega_2}.
\eea
In the case $\omega_1\gg\omega_2$ one has
\bea
G^{(4)}&&({\bf x}_3, {\bf B}_3,  z_3; {\bf x}_2, {\bf B}_2, z_2;\omega_1,\omega_2)\nn\\
&&\simeq G({\bf x}_3, {\bf x}_2, z_3 - z_2, \omega_1, \f{\sqrt{3}}{2}\Omega_1) G(\bar{{\bf B}}_3, \bar{{\bf B}}_2, z_3 - z_2, \omega_2, \Omega_2),\label{eq:G4soft}\\
\bar G^{(4)}&&({\bf x}_3, {\bf B}_3,  z_3; {\bf x}_2, {\bf B}_2, z_2;\omega_1,\omega_2)\nn\\
&&\simeq G({\bf x}_3, {\bf x}_2, z_3 - z_2, \omega_1, \f{\sqrt{3}}{2}\Omega_1) G^*(\bar{{\bf B}}_3, \bar{{\bf B}}_2, z_3 - z_2, \omega_2, \Omega_2),\label{eq:G4barsoft}
\eea
where
\be
\bar{{\bf B}}_2 \equiv {\bf B}_2 - \f{{\bf x}_2}{2}, \qquad\text{and}\qquad \bar{{\bf B}}_3 \equiv {\bf B}_3 - \f{{\bf x}_3}{2}.
\ee
\section{The double logarithmic correction to radiative energy loss}\label{sec:doublelog}

\begin{figure}
\begin{center}
\includegraphics[width=0.5\textwidth]{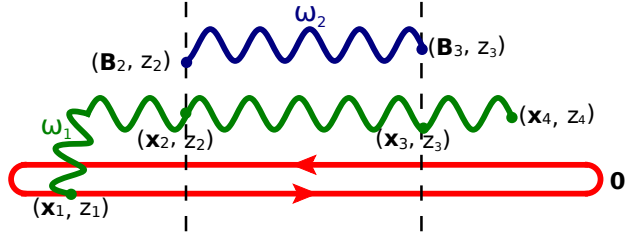}
\end{center}
\caption{Diagrams with fully-overlapping emissions. 13 diagrams can be constructed from this figure as follows. If the first gluon with energy $\omega_1$ is absorbed by the antiquark at $t=z_4$, the ends of the second gluon line with energy $\omega_2$ can be attached to the first gluon line or the diplole in 9 different ways. In contrast, the ends of the second gluon line can be attached to the first gluon line or the dipole in 4 different ways if the first gluon is absorbed by the quark. These 13 diagrams and their complex conjugates all contribute to $\Delta E_2$.}\label{fig:energyloss}
\end{figure}

In this section we calculate the double logarithmic correction to radiative energy loss. The double logarithmic correction comes from the diagrams with fully-overlapping emission defined in Sec. \ref{sec:BDMPSZ}. All these diagrams can be constructed either from Fig. \ref{fig:energyloss} or its complex conjugate. First, we give all the contributions of these diagrams to $\Delta E_2$ in a compact form in terms of $G^{(3)}$ and $G^{(4)}$ ($\bar G^{(4)}$) defined in the previous section. Next, we shall show that they give a double logarithmic correction to radiative energy loss, which is the same as that to the radiative $p_\perp$-broadening in Ref. \cite{Liou:Mueller:Wu:2013}. 
\subsection{Contributions from diagrams with fully-overlapping emission}
In total, there are 26 diagrams with fully-overlapping emission. Let us denote the energies of the two gluons respectively by $\omega_1$ and $\omega_2$ and  their formation times respectively by $t_1\equiv z_4-z_1$ and $t_2\equiv z_3-z_2$. And we assume that $\omega_1\gtrsim \omega_2$ and $t_1\gtrsim t_2$. As illustrated in Fig. \ref{fig:energyloss}, there are respectively 4 or 9 diagrams in which the gluon with energy $\omega_1$ is virtual or real. The 26 fully-overlapping emissions include these 13 diagrams and their complex conjugates. And one can obtain the corrections to the energy loss from fully-overlapping emissions by taking 2 times the real part of the contributions of these 13 diagrams.

There are some cancellations between the contributions from different diagrams. As a consequence of the conservation of probability, moving one gluon emission (absorption) vertex from the quark (antiquark) line to the antiquark (quark) line in a diagram only changes the overall sign of the contribution of the diagram\cite{Wu:2011}. To see such a cancellation easily, we use a diagramatic representation in which the integrations over all the time variables and transverse coordinates in Fig. \ref{fig:energyloss} are omitted. It is easy to see that we have the following cancellation\footnote{The other 5 diagrams, integrated over $\omega_1$ and $\omega_2$, are cancelled by those with two virtual gluons. This cancellation and that in (\ref{eq:Ia}) guarantee the conservation of probability.}
\bea\label{eq:Ia}
I_a=\int d\omega_1 d\omega_2&&\left\{\left[\left(\omega_1+\omega_2\right)\left(\begin{array}{c}\includegraphics[width=2.0cm]{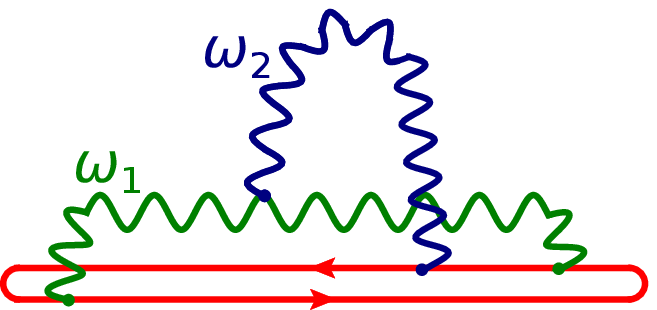}\end{array}\right)+
\omega_2\left(\begin{array}{c}\includegraphics[width=2.0cm]{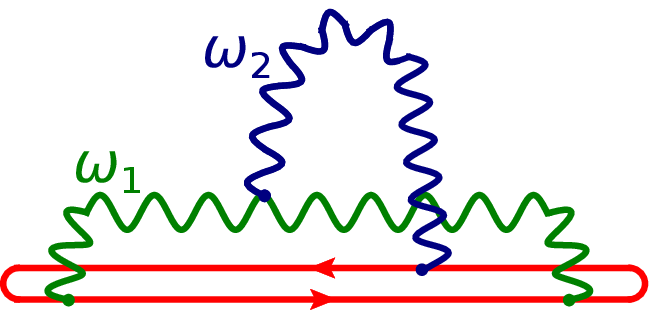}\end{array}\right)\right]\right.\nn\\
&&+\left[\left(\omega_1+\omega_2\right)\left(\begin{array}{c}\includegraphics[width=2.0cm]{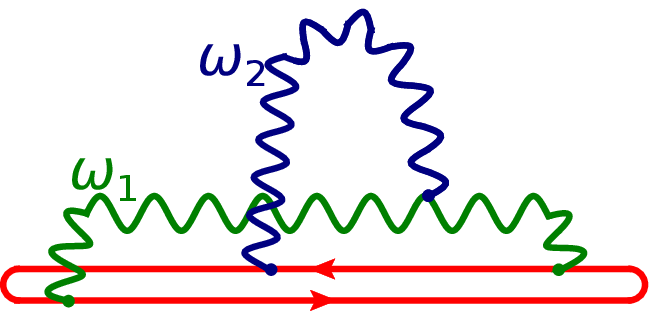}\end{array}\right)+
\omega_2\left(\begin{array}{c}\includegraphics[width=2.0cm]{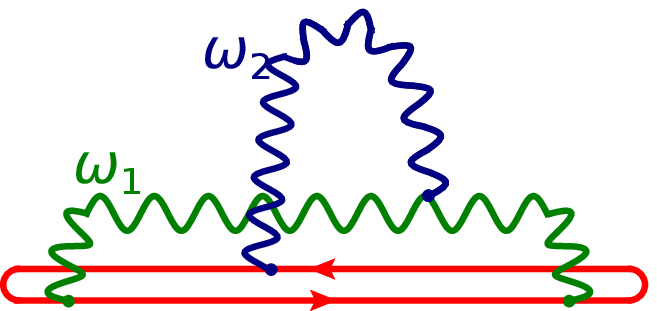}\end{array}\right)\right]\nn\\
&&+\left[\left(\omega_1+\omega_2\right)\left(\begin{array}{c}\includegraphics[width=2.0cm]{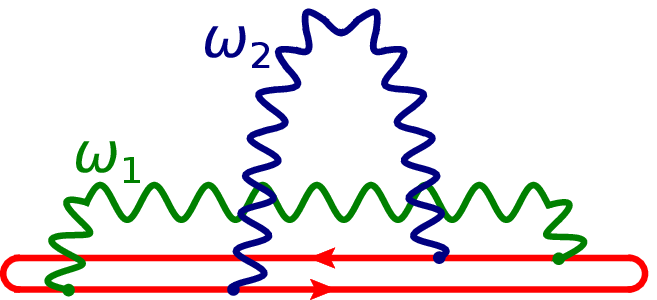}\end{array}\right)+
\omega_2\left(\begin{array}{c}\includegraphics[width=2.0cm]{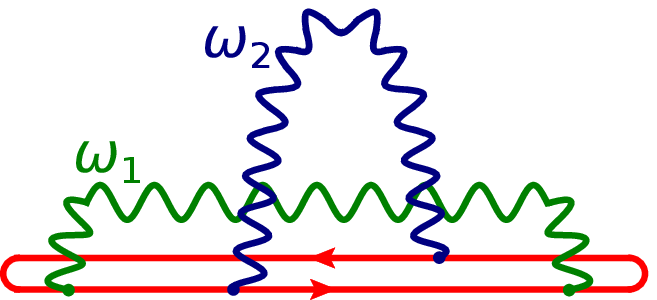}\end{array}\right)\right]\nn\\
&&+\left.\left[\left(\omega_1+\omega_2\right)\left(\begin{array}{c}\includegraphics[width=2.0cm]{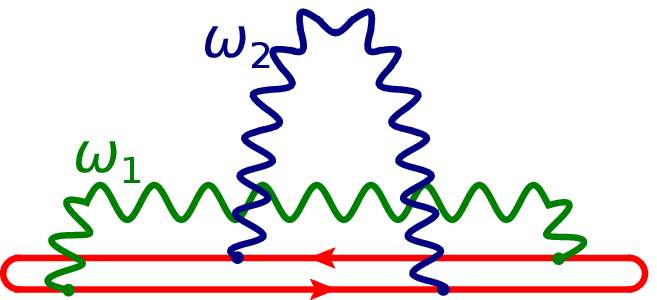}\end{array}\right)+
\omega_2\left(\begin{array}{c}\includegraphics[width=2.0cm]{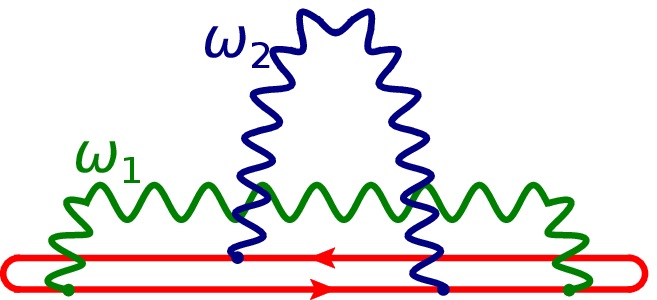}\end{array}\right)\right]\right\}\nn\\
=\int d\omega_1 d\omega_2&&~\omega_1~\left\{\left(\begin{array}{c}\includegraphics[width=2.0cm]{f04}\end{array}\right)
+\left(\begin{array}{c}\includegraphics[width=2.0cm]{f05}\end{array}\right)\right.\nn\\
&&+\left.\left(\begin{array}{c}\includegraphics[width=2.0cm]{f08}\end{array}\right)
+\left(\begin{array}{c}\includegraphics[width=2.0cm]{f09}\end{array}\right)\right\}.
\eea
Therefore, only these 9 diagrams with the gluon of energy $\omega_1$ being real contribute to $\Delta E_2$.

The contributions to $\Delta E_2$ from these 9 diagrams and their complex conjugates can be classified as follows
\bea
e_1\equiv&&\f{2}{(4\pi)^2}\text{Re}\int d\omega_1 d\omega_2\omega_1\left[\left(\begin{array}{c}\includegraphics[width=2.0cm]{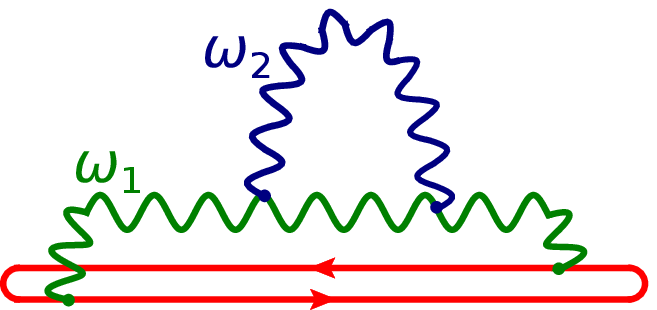}\end{array}\right)+\left(\begin{array}{c}\includegraphics[width=2.0cm]{f04}\end{array}\right)+\left(\begin{array}{c}\includegraphics[width=2.0cm]{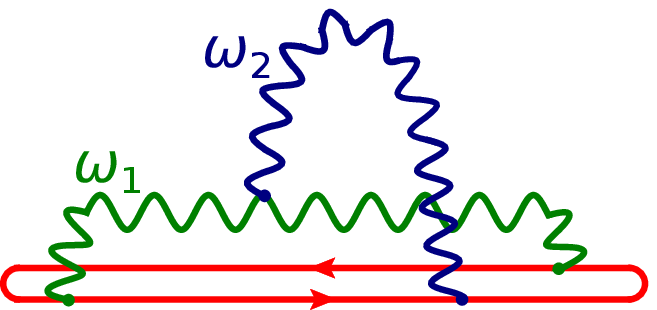}\end{array}\right)\right],\\
e_2\equiv&&\f{2}{(4\pi)^2}\text{Re}\int d\omega_1 d\omega_2\omega_1\left[\left(\begin{array}{c}\includegraphics[width=2.0cm]{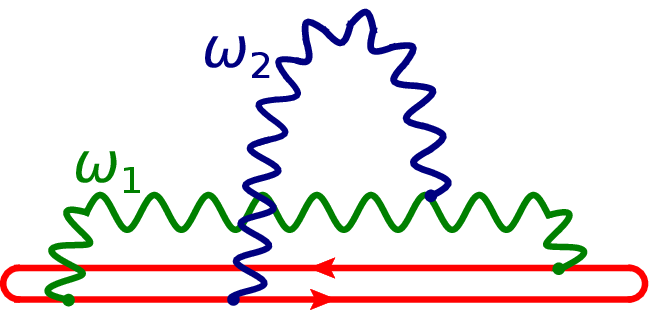}\end{array}\right)+\left(\begin{array}{c}\includegraphics[width=2.0cm]{f08}\end{array}\right)+\left(\begin{array}{c}\includegraphics[width=2.0cm]{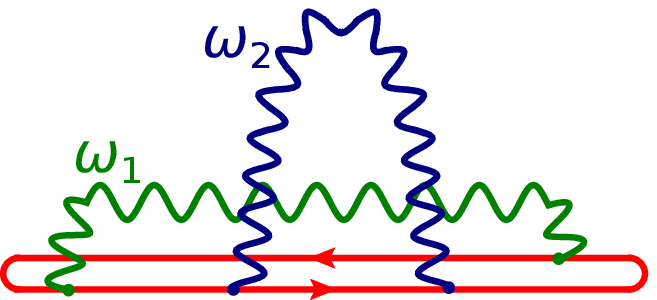}\end{array}\right)\right],\\
e_3\equiv&&\f{2}{(4\pi)^2}\text{Re}\int d\omega_1 d\omega_2\omega_1\left[\left(\begin{array}{c}\includegraphics[width=2.0cm]{f05}\end{array}\right)+\left(\begin{array}{c}\includegraphics[width=2.0cm]{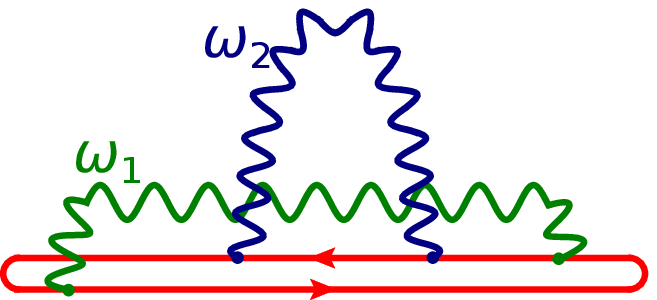}\end{array}\right)+\left(\begin{array}{c}\includegraphics[width=2.0cm]{f09}\end{array}\right)\right].
\label{eq:e3}
\eea
It is easy to show that the overall color factors for $e_1, e_2$ and $e_3$ are respectively given by $N_c C_R$, $N_c C_R/2$ and $N_c C_R/2$. In terms of $G^{(3)}$, $G^{(4)}$ and $\bar G^{(4)}$, we get, from the Feynman rules in Sec. \ref{sec:BDMPSZ}, the following compact expressions
\bea
e_1=2\alpha_s^2 N_c C_R~\text{Re}&&~\iiint \D_{{\bf x}_1}\cdot \left.\D_{{\bf x}_4}\left[ G^{(3)}({\bf x}_4, z_4, {\bf x}_3, z_3; \omega_1) G^{(3)}({\bf x}_2, z_2; {\bf x}_1, z_1; \omega_1)\right]\right|^{{\bf x}_1 =0 = {\bf x}_4}\nn\\
&&\times \D_{{\bf B}_2}\cdot \left.\D_{{\bf B}_3} G^{(4)}({\bf x}_3, {\bf B}_3, z_3; {\bf x}_2, {\bf B}_2 , z_2; \omega_1, \omega_2)\right|^{{\bf B}_2 = {\bf x}_2, {\bf B}_3 = 0}_{{\bf B}_2 = {\bf x}_2, {\bf B}_3 = {\bf x}_3},\label{eq:e1}\\
e_2=\alpha_s^2 N_c C_R~\text{Re}~&&\iiint \D_{{\bf x}_1}\cdot \left.\D_{{\bf x}_4}\left[ G^{(3)}({\bf x}_4, z_4, {\bf x}_3, z_3;\omega_1) G^{(3)}({\bf x}_2, z_2; {\bf x}_1, z_1;\omega_1)\right]\right|^{{\bf x}_1 =0 = {\bf x}_4}\nn\\
&&\times \D_{{\bf B}_2}\cdot \left.\D_{{\bf B}_3} G^{(4)}({\bf x}_3, {\bf B}_3, z_3; {\bf x}_2, {\bf B}_2 , z_2;\omega_1,\omega_2)\right|^{{\bf B}_2 = 0, {\bf B}_3 = {\bf x}_3}_{{\bf B}_2 = 0, {\bf B}_3 = 0},\label{eq:e2}\\
e_3=\alpha_s^2 N_c C_R~\text{Re}~&&\iiint \D_{{\bf x}_1}\cdot \left.\D_{{\bf x}_4}\left[ G^{(3)}({\bf x}_4, z_4, {\bf x}_3, z_3;\omega_1) G^{(3)}({\bf x}_2, z_2; {\bf x}_1, z_1;\omega_1)\right]\right|^{{\bf x}_1 =0 = {\bf x}_4}\nn\\
&&\times \D_{{\bf B}_2}\cdot \left.\D_{{\bf B}_3} \bar G^{(4)}({\bf x}_3, {\bf B}_3, z_3; {\bf x}_2, {\bf B}_2 , z_2;\omega_1,\omega_2)\right|^{{\bf B}_2 = 0, {\bf B}_3 = {\bf x}_3}_{{\bf B}_2 = 0, {\bf B}_3 = 0},\label{eq:e3}
\eea
where the short-hand notation
\be
\iiint\equiv\int \f{d\omega_1}{\omega_1^2}\f{d\omega_2}{\omega_2^3}\int_0^L dz_4 \int_0^{z_4} dz_3 \int_0^{z_3} dz_2 \int_0^{z_2} dz_1 \int d^2{\bf x}_2 d^2 {\bf x}_3.
\ee 
And the contributions from all the fully-overlapping emissions to the energy loss are given by
\bea
\Delta E_2 =&& e_1 + e_2 + e_3.\label{eq:DEdb}
\eea
It is still too complicated to be evaluated analytically even in the harmonic oscillator approximation. In the next subsection, we shall evaluate it in the double logarithmic approximation following Ref. \cite{Liou:Mueller:Wu:2013}.

\subsection{$\Delta E_2$ in the double logarithmic approximation}
The calculation of $\Delta E_2$ in eq. (\ref{eq:DEdb}) is simplified in the double logarithmic region. The first gluon with energy $\omega_1$, similar to that in the case with one-gluon emission, typically has $t_1\equiv z_4 - z_1 \simeq L$, $\omega_1\simeq \hat{q} L^2$ and $|{\bf x}|^2 \simeq \f{1}{\hat{q} L}$. In the double logarithmic region\cite{Liou:Mueller:Wu:2013}, the second gluon with energy $\omega_2$ typically has  
\be
|{\bf B}|^2 \gtrsim |{\bf x}|^2,\qquad \omega_2 \lesssim \omega_1, \qquad\text{and}\qquad t_2\lesssim \sqrt{\f{\omega_2}{\hat{q}_A}}\lesssim\sqrt{\f{\omega_1}{\hat{q}_A}}\simeq t_1.\ee 
In this region one can use the following approximation
\bea
&&G^{(4)}({\bf x}_3, {\bf B}_3,  z_3; {\bf x}_2, {\bf B}_2, z_2;\omega_1, \omega_2)\simeq \delta({\bf x}_3-{\bf x}_2)e^{-\f{3}{16} \hat{q}_A x_2^2 t_2} G(\bar {\bf B}_3, \bar {\bf B}_2, t_2, \omega_2, \Omega_2),\label{eq:G4db}\\
&&\bar G^{(4)}({\bf x}_3, {\bf B}_3,  z_3; {\bf x}_2, {\bf B}_2, z_2;\omega_1, \omega_2)\simeq \delta({\bf x}_3-{\bf x}_2)e^{-\f{3}{16} \hat{q}_A x_2^2 t_2} G^*(\bar {\bf B}_3, \bar {\bf B}_2, t_2, \omega_2, \Omega_2).\label{eq:G4bardb}
\eea
By inserting (\ref{eq:G3sp}), (\ref{eq:G4db}) and (\ref{eq:G4bardb}) into (\ref{eq:DEdb}), we have
\bea
\Delta E_2 \simeq&& 2\alpha_s^2 N_c C_R\int \f{d\omega_1}{\omega_1^2}\f{d\omega_2}{\omega_2^3}\int_0^L dz_4 \int_0^{L-z_4} dz_1 \int_{z_1}^{z_4} dz_3 \int_{0}^{z_3-z_1} dt_2 \int d^2{\bf x}_2 e^{-\f{3}{16} \hat{q}_A x_2^2 t_2}\nn\\
&&\times\text{Re}\left\{\D_{{\bf x}_1}\cdot \left.\D_{{\bf x}_4}\left[ G({\bf x}_4, {\bf x}_2, z_4 - z_3;\omega_1, \Omega_1) G({\bf x}_2, {\bf x}_1, z_3-t_2 - z_1;\omega_1, \Omega_1)\right]\right|^{{\bf x}_1 =0 = {\bf x}_4}\right.\nn\\
&&\times\left.\left[\D_{{\bf B}_2}\cdot \left.\D_{{\bf B}_3} G({\bf B}_3-\f{{\bf x}_2}{2}, {\bf B}_2-\f{{\bf x}_2}{2} , t_2;\omega_2, \Omega_2)\right|^{{\bf B}_2 = {\bf x}_2, {\bf B}_3 = 0}_{{\bf B}_2 = {\bf x}_2, {\bf B}_3 = {\bf x}_2}\right.\right.\nn\\
&&+\left.\left.\text{Re}\D_{{\bf B}_2}\cdot \left.\D_{{\bf B}_3} G({\bf B}_3-\f{{\bf x}_2}{2}, {\bf B}_2-\f{{\bf x}_2}{2} , t_2;\omega_2, \Omega_2)\right|^{{\bf B}_2 = 0, {\bf B}_3 = {\bf x}_2}_{{\bf B}_2 = 0, {\bf B}_3 = 0}\right]
\right\}.
\eea
In the double logarithmic region one has $t_1\lesssim t_2$. Therefore, we can neglect the difference between $z_3$ and $z_2$ and write
\bea\label{eq:DErad}
\Delta E_2 \simeq && 2\alpha_s N_c\int \f{d\omega_1}{\omega_1^2}\int_0^L dz_4 \int_0^{L-z_4} dz_1 \int_{z_1}^{z_4} dz_3  \int d^2{\bf x}_2 \text{Re}\left\{S({\bf x}_2,z_3 - z_1)\right.\nn\\
&&\times\left.\left.\D_{{\bf x}_1}\cdot \D_{{\bf x}_4}\left[ G({\bf x}_4, {\bf x}_2,z_4 - z_3;\omega_1, \Omega_1) G({\bf x}_2, {\bf x}_1, z_3-z_1;\omega_1, \Omega_1)\right]\right|^{{\bf x}_1 =0 = {\bf x}_4}\right\},
\eea
where
\bea
S({\bf x}_2, &&z_3 - z_1)\equiv \alpha_s C_R \int\f{d\omega_2}{\omega_2^3}\int_{0}^{z_3-z_1} dt_2e^{-\f{3}{16} \hat{q}_A x_2^2 t_2}\nn\\
&&\times\left.\left[\D_{{\bf B}_2}\cdot \left.\D_{{\bf B}_3} G({\bf B}_3-\f{{\bf x}_2}{2}, {\bf B}_2-\f{{\bf x}_2}{2} , t_2;\omega_2, \Omega_2)\right|^{{\bf B}_2 = {\bf x}_2, {\bf B}_3 = 0}_{{\bf B}_2 = {\bf x}_2, {\bf B}_3 = {\bf x}_2}\right.\right.\nn\\
&&+\left.\text{Re}\D_{{\bf B}_2}\cdot \left.\D_{{\bf B}_3} G({\bf B}_3-\f{{\bf x}_2}{2}, {\bf B}_2-\f{{\bf x}_2}{2} , t_2;\omega_2, \Omega_2)\right|^{{\bf B}_2 = 0, {\bf B}_3 = {\bf x}_2}_{{\bf B}_2 = 0, {\bf B}_3 = 0}\right].
\eea

\begin{figure}
\begin{center}
\includegraphics[width=0.5\textwidth]{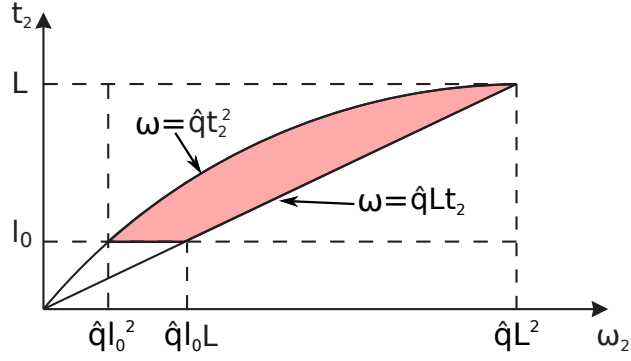}
\end{center}
\caption{Kinetic region for double logarithmic correction\cite{Liou:Mueller:Wu:2013}. The double logarithmic region is enclosed by the following three curves: $t_2 = l_0$, $\omega_2 = \hat{q} L t_2$ and $\omega_2 = \hat{q} t_2^2$.
}\label{fig:logregion}
\end{figure}

Now we are ready to show how the double logarithmic correction to radiative $p_\perp$-broadening shows up in the calculation of radiative energy loss. By dropping terms independent of $\hat{q}$\footnote{In this way one calculates the medium-induced energy loss as explained in Refs. \cite{BDMPS:1996,BDMPS:1998,Zakharov:1996,Zakharov:1997}. A consequence of such a subtraction is that one drops the vacuum diagrams (including the UV divergent ones) and, therefore, ignores the effects of running coupling. The consequences of running coupling to $\bra p_\perp^2\ket$ or the renormalized $\hat{q}$ is studied in \cite{ Iancu:Triantafyllopoulos:2014}, which is beyond the scope of this paper.} and keeping only the double logarithmic terms proportional to $x_2^2$, we have
\bea\label{eq:Slog}
S({\bf x}_2, &&z_3 - z_1)\simeq-\f{\alpha_s N_c}{4 \pi}\hat{q} x_2^2\int\f{d\omega_2}{\omega_2} \int_{0}^{z_3-z_1} \f{dt_2}{ t_2 }=-\f{\alpha_s N_c}{32 \pi}\hat{q} x_2^2\ln^2\left(\f{z_3-z_1}{l_0}\right)^2,
\eea
where $t_2$ and $\omega_2$ have been integrated over the double logarithmic region in Fig. \ref{fig:logregion} with $L$ replaced by $z_3-z_1$. As we shall show below, one can simply replace $z_3 - z_1$ in (\ref{eq:Slog}) by $L$ in the double logarithmic approximation. Therefore, such a double logarithmic result is exactly the same as that of radiative $\bra p_\perp^2 \ket$ and eq. (\ref{eq:Slog}) is the same as eq. (26) in Ref. \cite{Liou:Mueller:Wu:2013} divided by $L$. 

Let us evaluate the double logarithmic correction to radiative energy loss. Inserting (\ref{eq:Slog}) into (\ref{eq:DErad}) and integrating out $x_2$ gives
\bea
\Delta E_2 \simeq &&  \f{\alpha_s^2 N_c^2}{4 \pi^2}\hat{q}\text{Re}~i\int \f{d\omega_1}{\omega_1}\int_0^L dz_4 \int_0^{L-z_4} dz_1 \nn\\
&&\times\int_{z_1}^{z_4} dz_3 \ln^2\left(\f{z_3-z_1}{l_0}\right)^2\f{\Omega_1 \sin(\Omega_1 (z_4 - z_3)) \sin(\Omega_1 (z_3 - z_1))}{\sin^3(\Omega_1 (z_4 - z_1))}.
\eea
Since the integrand on the right-hand side of the above equation is proportional to $z_3 - z_1$ as $z_3\to z_1$, the leading double logarithmic term from the integration over $z_3$ can be obtained simply by integration by parts, that is,  
\bea
\Delta E_2 \simeq && \f{\alpha_s^2 N_c^2}{4 \pi^2}\hat{q}\text{Re}~i \int \f{d\omega_1}{\omega_1}\int_0^L dz_4 \int_0^{L-z_4} dz_1 \ln^2\left(\f{z_4-z_1}{l_0}\right)^2 \nn\\
&&\times \int_{z_1}^{z_4} dz_3\f{\Omega_1 \sin(\Omega_1 (z_4 - z_3)) \sin(\Omega_1 (z_3 - z_1))}{\sin^3(\Omega_1 (z_4 - z_1))}\nn\\
=&& \f{\alpha_s^2 N_c^2}{8 \pi^2}\hat{q}\text{Re}~i \int \f{d\omega_1}{\omega_1}\int_0^L dt_1 (L-t_1) \ln^2\left(\f{t_1}{l_0}\right)^2 \f{\Omega_1 \sin(\Omega_1 t_1) - \Omega_1 t_1 \cos(\Omega_1 t_1)}{\sin^3(\Omega_1 t_1)}.
\eea
Similarly, the leading double logarithmic term from the integration over $t_1$ is of the form
\bea\label{eq:DE2log}
\Delta E_2 \simeq && \f{\alpha_s^2 N_c^2}{8 \pi^2}\hat{q} \ln^2\left(\f{L}{l_0}\right)^2\text{Re}~i \int_0^\infty \f{d\omega_1}{\omega_1}\int_0^L dt_1 (L-t_1)  \f{\Omega_1 \sin(\Omega_1 t_1) - \Omega_1 t_1 \cos(\Omega_1 t_1)}{\sin^3(\Omega_1 t_1)}\nn\\
=&&\f{\alpha_s^2 N_c^2}{16 \pi^2}\hat{q} \ln^2\left(\f{L}{l_0}\right)^2\text{Re}~i \int_0^\infty \f{d\omega_1}{\omega_1}\f{1}{\Omega_1^2}\left(1-\frac{\Omega_1 L}{ \tan (\Omega_1 L)}\right)\nn\\
=&&\f{\alpha_s^2 N_c^2}{16 \pi^2}\hat{q}L^2\ln^2\left(\f{L}{l_0}\right)^2\text{Re}\int_0^\infty \f{d\hat{\omega}}{\hat{\omega}^3}\left(\f{(1-i)\hat{\omega}}{\tan((1-i)\hat{\omega})}-1\right)\nn\\
=&&\f{\alpha_s N_c}{12}L\times\f{\alpha_s N_c}{8 \pi}\hat{q} \ln^2\left(\f{L}{l_0}\right)^2L=\f{\alpha_s N_c}{12}L\bra p_\perp^2\ket_{rad},
\eea
where $\hat{\omega}\equiv \f{1}{2}\sqrt{\f{\hat{q}_A L^2}{\omega}}$ and recall that radiative energy loss due to one-gluon emission is given by\cite{BDMPS:1998}
\bea
\Delta E_1 = && 2\alpha_s C_R\text{Re}\int \f{d\omega_1}{\omega_1^2}\int_0^L dt(L-t) \D_{{\bf x}_1}\cdot \D_{{\bf x}_4}\left.\left[ G({\bf x}_4, {\bf x}_1,t;\omega_1, \Omega_1)\right.\right.\nn\\
&&\left.\left.-G({\bf x}_4, {\bf x}_1,t;\omega_1, 0)\right]\right|^{{\bf x}_1 =0 = {\bf x}_4}=\f{\alpha_s N_c}{12} \hat{q} L^2.
\eea

The resummation of the double logarithmic correction in eq. (\ref{eq:DE2log}) can be carried out in exactly the same way as that in the calculation of $\bra p_\perp^2\ket$. For $(n+1)$-gluon emission, one has
\bea\label{eq:SlogResum}
S({\bf x}_2, &&z_3 - z_1)\simeq-\f{1}{4} \hat{q} x_2^2 \f{1}{n!(n+1)!}\left[\f{\alpha_s N_c}{4 \pi}\ln^{2}\left(\f{z_3-z_1}{l_0}\right)^2\right]^n,
\eea
which gives
\bea\label{eq:DE2logResum}
\Delta E_{n+1}=&&\f{\alpha_s N_c}{12}L\f{\hat{q} L}{n!(n+1)!}\left[\f{\alpha_s N_c}{4 \pi}\ln^{2}\f{L^2}{l_0^2}\right]^n.\eea
Therefore, the total energy loss is given by
\bea
\Delta E = \sum\limits_{n=0}^\infty\Delta E_{n+1} = \f{\alpha_s N_c}{12}L\f{\hat{q}L}{\sqrt{\f{\alpha_s N_c}{4 \pi}}\ln\f{L^2}{l_0^2}}I_1\left[\sqrt{\f{\alpha_s N_c}{\pi}}\ln\f{L^2}{l_0^2} \right] = \f{\alpha_s N_c}{12}L\bra p_\perp^2\ket.
\eea
%
%
%
%
\section{Conclusions}\label{sec:conclusion}
In this paper we calculate the double logarithmic correction to radiative energy loss of a high-energy parton in the generalized BMDMPS-Z formalism\cite{Wu:2011,Liou:Mueller:Wu:2013}. Radiative energy loss per unit length due to one-gluon emission is given by\cite{BDMPS:1998}  
\be
-\f{dE_1 }{dz} = \f{\alpha_s N_c}{12} \hat{q} L.
\ee
In this case it is dominated by radiating one gluon with the maximum energy $\omega_c \simeq \hat{q} L^2$ and the formation time $t_c\simeq L$. And within $t_c$ the gluon is of a typical size $x^2 \sim\f{1}{\hat{q} L}$. The double logarithmic correction comes from the diagrams by adding a second gluon with the two emission times both lie in between those of the first gluon of energy $\omega_c$. In the kinetic region for the double logarithmic correction this second gluon has a smaller energy, a shorter formation time and a larger size than the first gluon. It modifies the transverse momentum broadening of the first gluon and, therefore, contributes to the radiative energy loss according to the parametric estimate in eq. (\ref{eq:dept2}). Our detailed calculation confirms this picture and we find that the double logarithmic correction to the energy loss due to two-gluon emission satisfies
\be
-\f{dE_2 }{dz} = \f{\alpha_s N_c}{12}\bra p_\perp^2 \ket_{rad},
\ee 
where $\bra p_\perp^2 \ket_{rad}$ is given in eq. (\ref{eq:pt2rad}), which is obtained in Ref. \cite{Liou:Mueller:Wu:2013}. Moreover, the resummation  of the double logarithmic terms can be carried out to give
\be
-\f{dE }{dz} = \f{\alpha_s N_c}{12}\bra p_\perp^2 \ket,
\ee 
where $\bra p_\perp^2\ket$ is given in eq. (\ref{eq:pt2}). Our result agrees with that by using the renormalized $\hat{q}$ in Refs. \cite{Blaizot:Mehtar-Tani:2014, Iancu:2014}.
\section*{Acknowledgements}
The author would like to thank A. H. Mueller, F. Dominguez, Y. Mehtar-Tani and E. Iancu for inspiring discussions and/or suggestions. This work is supported by the Agence Nationale de la Recherche project \# 11-BS04-015-01.


\begin{thebibliography}{99}

\bibitem{Adler:2003} 
  S.~S.~Adler {\it et al.}  [PHENIX Collaboration],
  Phys.\ Rev.\ Lett.\  {\bf 91}, 072301 (2003)
  [nucl-ex/0304022].

\bibitem{Adams:2003} 
  J.~Adams {\it et al.}  [STAR Collaboration],
  Phys.\ Rev.\ Lett.\  {\bf 91}, 172302 (2003)
  [nucl-ex/0305015].

\bibitem{ALICE:2010} 
  K.~Aamodt {\it et al.}  [ALICE Collaboration],
  Phys.\ Lett.\ B {\bf 696}, 30 (2011)
  [arXiv:1012.1004 [nucl-ex]].

\bibitem{CMS:2012} 
  S.~Chatrchyan {\it et al.}  [CMS Collaboration],
  Eur.\ Phys.\ J.\ C {\bf 72}, 1945 (2012)
  [arXiv:1202.2554 [nucl-ex]].


\bibitem{ATLAS:2012} 
  G.~Aad {\it et al.}  [ATLAS Collaboration],
  Phys.\ Lett.\ B {\bf 719}, 220 (2013)
  [arXiv:1208.1967 [hep-ex]].


\bibitem{Muller:2012} 
  B.~Muller, J.~Schukraft and B.~Wyslouch,
  Ann.\ Rev.\ Nucl.\ Part.\ Sci.\  {\bf 62}, 361 (2012)
  [arXiv:1202.3233 [hep-ex]].

\bibitem{BDMPS:1996pt} 
  R.~Baier, Y.~L.~Dokshitzer, A.~H.~Mueller, S.~Peigne and D.~Schiff,
  Nucl.\ Phys.\ B {\bf 484}, 265 (1997)
  [hep-ph/9608322].

\bibitem{BDMPS:1996}
  R.~Baier, Y.~L.~Dokshitzer, A.~H.~Mueller, S.~Peigne and D.~Schiff,
  Nucl.\ Phys.\  B {\bf 483}, 291 (1997)
  [arXiv:hep-ph/9607355].
    
\bibitem{BDMPS:1998}
  R.~Baier, Y.~L.~Dokshitzer, A.~H.~Mueller and D.~Schiff,
  Nucl.\ Phys.\  B {\bf 531}, 403 (1998)
  [arXiv:hep-ph/9804212].

\bibitem{Wu:2011} 
  B.~Wu,
  JHEP {\bf 1110}, 029 (2011)
  [arXiv:1102.0388 [hep-ph]].


\bibitem{Liou:Mueller:Wu:2013} 
  T.~Liou, A.~H.~Mueller and B.~Wu,
  Nucl.\ Phys.\ A {\bf 916}, 102 (2013)
  [arXiv:1304.7677 [hep-ph]].

\bibitem{Panero:2013} 
  M.~Panero, K.~Rummukainen and A.~Schäfer,
  Phys.\ Rev.\ Lett.\  {\bf 112}, 162001 (2014)
  [arXiv:1307.5850 [hep-ph]].

    
\bibitem{CasalderreySolana:Wang:2007} 
  J.~Casalderrey-Solana and X.~N.~Wang,
  Phys.\ Rev.\ C {\bf 77}, 024902 (2008)
  [arXiv:0705.1352 [hep-ph]].

\bibitem{Xing:2014} 
  H.~Xing, Z.~B.~Kang, E.~Wang and X.~N.~Wang,
  arXiv:1407.8506 [hep-ph].

\bibitem{Blaizot:Mehtar-Tani:2014} 
  J.~-P.~Blaizot and Y.~Mehtar-Tani,
  arXiv:1403.2323 [hep-ph].

\bibitem{Iancu:2014} 
  E.~Iancu,
  arXiv:1403.1996 [hep-ph].


  \bibitem{Zakharov:1996}
  B.~G.~Zakharov,
  JETP Lett.\  {\bf 63} (1996) 952
  [arXiv:hep-ph/9607440].
  
  \bibitem{Zakharov:1997}
  B.~G.~Zakharov,
  JETP Lett.\  {\bf 65} (1997) 615
  [arXiv:hep-ph/9704255].

\bibitem{Kovner:Wiedemann:2003} 
  A.~Kovner and U.~A.~Wiedemann,
  In *Hwa, R.C. (ed.) et al.: Quark gluon plasma* 192-248
  [hep-ph/0304151].

\bibitem{Majumder:2014} 
  A.~Majumder,
  arXiv:1405.2019 [nucl-th].

\bibitem{BDMS:2001:uncorrelated} 
  R.~Baier, Y.~L.~Dokshitzer, A.~H.~Mueller and D.~Schiff,
  JHEP {\bf 0109}, 033 (2001)
  [hep-ph/0106347].

\bibitem{Iancu:Triantafyllopoulos:2014} 
  E.~Iancu and D.~N.~Triantafyllopoulos,
  arXiv:1405.3525 [hep-ph].

  
\end{thebibliography}
\end{document}